\documentclass[twocolumn,showpacs,preprintnumbers,amsmath,amssymb]{revtex4}

\newcommand{\la}{\langle}
\newcommand{\ra}{\rangle}
\newcommand{\be}{\begin{equation}}
\newcommand{\ee}{\end{equation}}
\newcommand{\bea}{\begin{eqnarray}}
\newcommand{\eea}{\end{eqnarray}}

\usepackage{graphicx}
\usepackage{dcolumn}
\usepackage{bm}
\usepackage{epsfig}

\begin{document}

\title{Nonequilibrium thermodynamics of a squeezed harmonic oscillator}

\author{Fernando  Galve and Eric Lutz}
 \affiliation{Department of Physics, University of Augsburg, 86135 Augsburg, Germany}
\date{\today}

\begin{abstract}
We consider the thermodynamic properties of the  squeezed vacuum state of a frequency--modulated quantum harmonic oscillator. We analytically relate the squeezing parameter to the irreversible work and the degree of nonadiabaticity of the frequency transformation. We furthermore determine the optimal modulation that leads to maximal squeezing, and discuss its implementation as well as the detection of squeezing  in single cold ion traps.
\end{abstract}

\pacs{42.50.-p, 03.65.-w }

\maketitle

The time--dependent quantum harmonic oscillator serves as a model system for a variety of physical problems. Notable examples include the generation of nonclassical states \cite{aga91} and the dynamics of cold   ions in Paul traps \cite{wineland}.  More recently, it has  played a major role in the study  of cosmological particle creation \cite{sch07} and nonequilibrium quantum thermodynamics \cite{hub08}. An experimental investigation of a harmonic atom trap with frequency jumps has   moreover been reported in Ref.~\cite{ahm06}. The importance of the time--dependent quantum oscillator stems from the fact that it is an exactly solvable system. The propagator and the transitions probabilities of an oscillator with time--dependent frequency and time--dependent linear driving have been first derived by Husimi  using a Gaussian wave function  ansatz  \cite{Husimi}. A more general method based on invariant operators has been later developed by Lewis and Riesenfeld \cite{lew69}.

A remarkable property is that the state of the oscillator is squeezed when its angular frequency is changed in a nonadiabatic way \cite{jan87,gra87}. The degree of squeezing  depends on  the specific frequency modulation considered and various protocols have been discussed \cite{ma89,lo90,jan92,ave94}.
Efficient schemes for the production of squeezed states are nowadays essential  for high precision measurements,   such as in interferometric gravity wave detectors \cite{ligo}. In addition,  squeezing appears as a crucial resource for entanglement creation  in quantum information applications with continuous variables \cite{kra03,gal08}. The determination of an optimal squeezing protocol is thus of high importance.

The quantum harmonic oscillator with arbitrary frequency modulation has  recently been studied from a thermodynamical point of view in Ref.~\cite{work}. Using the general formula for the probability density of quantum work introduced in Ref.~\cite{tal07}, the statistics of the total  work performed on an isolated, but initially thermal,  oscillator during adiabatic and nonadiabatic variations of its angular frequency has been determined. In particular, an expression for the irreversible work, also often referred to as dissipative work \cite{rit04}, has been obtained. The irreversible work is defined as the difference between the total work $\la W\ra$ and the free energy difference $\Delta F$ (the reversible work): $\la W_{irr} \ra = \la W\ra - \Delta F$. The total work is here given by the difference between final and initial energies of the oscillator, $\la W\ra = \la H(\tau)\ra - \la H(0) \ra$, where $H(t)$ is the time--dependent Hamilton operator the quantum oscillator. On the other hand, the free energy difference can be written in  terms of the partition function $Z(t)$ in the usual manner as  $\Delta F=F(\tau) - F(0)=-kT\ln Z(\tau)/Z(0)$,  with $T$ the initial temperature of the oscillator. The irreversible work corresponds to the energy absorbed by the system during a nonadiabatic transformation and is therefore zero for an adiabatic, reversible, change of the frequency. It also represents the energy that  would be dissipated into a heat bath at temperature $T$, were the system weakly coupled to one \cite{kaw07}. The irreversible work has been related to  a parameter  $Q^*$ (see Eq.~\eqref{10} below), originally introduced by Husimi. The latter can be regarded as a measure of the degree of nonadiabaticity  \cite{work}: $Q^*$ is  unity for an adiabatic transformation of the oscillator's frequency and increases monotonically the less adiabatic a transformation is. 

In this paper,  we express the amount of squeezing generated by an arbitrary frequency change in terms of the nonadiabaticity parameter  $Q^*$ and the irreversible work $\la W_{irr} \ra$ for an oscillator initially in the ground state. By considering the thermodynamics of vacuum squeezing and, in particular, by quantifying the degree of  squeezing with the help of  the nonadiabaticity parameter, we are  able to
extend initial studies of squeezing production in frequency--modulated harmonic oscillators. We further use optimal control theory \cite{oct} to determine the modulation that leads to maximal squeezing and compare the results with those obtained in Ref.~\cite{jan92}.  We finally discuss the experimental implementation of the optimal modulation as well as the determination of the thermodynamic and squeezing properties of the quantum oscillator using  single ions in linear Paul traps.   

{\it Thermodynamics of squeezing.}
A quantum harmonic oscillator with time--dependent frequency $\omega(t)$ and unit mass is described by
the Hamilton operator, 
\begin{equation}
\label{1}
H(t) =\frac{1}{2}\left(p^2+\omega^2(t) q^2\right) \ .
\end{equation}
We consider a protocol where the frequency is changed from an initial value $\omega(0)=\omega_0$ to a final value $\omega(\tau)=\omega_1$ during time $\tau$. The solution of the Heisenberg equations of motion for the position and momentum operators $q(t)$ and $p(t)$ can then be written in the general form,
\bea
q(t)&=&q(0) Y(t)+ p(0) X(t) \ , \\
p(t)&=&\dot q(t) \ ,
\eea
where the two functions $X(t)$ and $Y(t)$ are the solutions of the equation of motion of  the corresponding {\it classical} oscillator, $\ddot X + \omega^2(t) X=0$,  with  initial conditions,  $X(0)=0$, $\dot{X}(0)=1$ and $Y(0)=1$, $\dot{Y}(0)=0$. The latter ensure that the canonical commutation relation, $[x,p]=i$, is satisfied (we set $\hbar =1$ throughout). 
The position and momentum variances for an oscillator initially in the ground state follow accordingly as,
\begin{eqnarray}
\la q^2 \ra &=&\frac{Y^2}{2\omega_0}+\frac{\omega_0X^2}{2} \label{eq4}\ ,\\
\la p^2\ra &=&\frac{\dot{Y}^2}{2\omega_0}+\frac{\omega_0\dot{X}^2}{2}\label{eq5} \ ,\\
\la q p \ra &=&\frac{Y\dot{Y}}{2\omega_0}+\frac{\omega_0X\dot{X}}{2} \ .
\end{eqnarray}
As shown by Husimi, details about a specific protocol $\omega(t)$ is fully contained in a function $Q^*(t)$  defined as \cite{Husimi}:
\begin{equation}
Q^*(t)=\frac{1}{2\omega_0\omega(t)}\left(\omega_0^2(\omega^2(t)X^2+\dot{X}^2)+\omega^2(t) Y^2+\dot{Y}^2     \right)
\end{equation}
By  using Eqs.~\eqref{eq4} and \eqref{eq5}, we easily find that
\begin{eqnarray}
Q^*(t)&=&\omega(t)\left(\frac{Y^2}{2\omega_0}+\frac{\omega_0X^2}{2}\right)+\frac{1}{\omega(t)}\left(\frac{\dot{Y}^2}{2\omega_0}+\frac{\omega_0\dot{X}^2}{2}\right) \nonumber \\
&=&\frac{1}{\omega(t)}\left(\la p^2\ra+\omega^2(t)\la q^2\ra\right) \label{8} \ ,
\end{eqnarray}
where the parenthesis is recognized as twice the expectation value of Hamiltonian \eqref{1}. 
We thus obtain that the mean  energy of the oscillator at any given time is simply
\be 
\la H(t) \ra=\frac{\omega(t)}{2}\, Q^*(t) \ .
\label{eq9}
\ee
The above equation  provides some insight into the physical meaning of the parameter $Q^*(t)$: For an adiabatic transformation,  $Q^*(t)=1$, and the mean energy of the oscillator is just given by the ground state energy. This corresponds to the known classical result that the action of the system, defined as the ratio of the energy and the angular frequency, is a time--independent constant. On the other hand, for a nonadiabatic change of the frequency, the value of $Q^*(t)>1$ indicates how far the final state of the oscillator is from its initial (equilibrium) ground state. The latter statement can be made more precise by directly computing the irreversible work $\left<W_{irr}\right>$ at the final time $\tau$. By evaluating the total work $\la W\ra$ and the free energy difference $\Delta F$ for Hamiltonian \eqref{1}, one finds \cite{work},
\begin{equation}
\label{10}
\left<W_{irr}\right>=\frac{\omega(\tau)}{2}\left(Q^*(\tau)-1\right)\ .
\end{equation}
The irreversible work is therefore zero for adiabatic transformations, as expected, and grows linearly with the nonadiabaticity parameter $Q^*(t)$.

Let us now establish a relationship between the irreversible work and the degree of squeezing of the harmonic oscillator. At any given time, a squeezed oscillator state can be parameterized as \cite{lew},
\begin{eqnarray}
\left< x^2\right> &=&\frac{1}{2\omega}\left(e^{-2r}\text{cos}^2\theta+e^{2r}\text{sin}^2\theta\right) \label{11}\ ,\\
\left<p^2\right>&=&\frac{\omega}{2}\left(e^{-2r}\text{sin}^2\theta+e^{2r}\text{cos}^2\theta\right) \label{12}\ , \\
\la qp \ra&=&\text{sinh}(2r) \sin \theta \cos \theta\ .
\end{eqnarray}
The time dependence of the squeezing parameter $r(t)$  and the rotation angle $\theta(t)$ is controlled by the frequency modulation $\omega(t)$.  One should note that the mean values $\left<x\right>$ and $\left<p\right>$ remain here  zero at all times.
By inserting Eqs.~\eqref{11}--\eqref{12} into Eq.~\eqref{8}, we obtain the following relation between  $Q^*$ and the squeezing parameter $r$,
\begin{equation}
\label{14}
Q^*(t) =\text{cosh}2r(t)\ .
\end{equation}
Equation \eqref{14} is an important result that directly connects  the degree of squeezing of the oscillator  to the nonadiabaticity  parameter, and thus to the frequency modulation $\omega(t)$.  It clearly shows that squeezing requires a nonadiabatic change of the frequency, $Q^*(t)>1$, and that large squeezing implies an exponential increase of $Q^*(t)$: $Q^*(t) \sim \exp(2 r(t))/2$. By further combining Eqs.~\eqref{10} and \eqref{14}, we arrive at
\begin{equation}\label{15}
\left<W_{irr}\right>=\omega(\tau)  \sinh^2 r(\tau) \ .
\end{equation}
Two limiting cases follow from this equation: $\left<W_{irr}\right>\sim\omega\ r^2$ for small squeezing and $\left<W_{irr}\right>\sim\omega\ e^{2r}/4$ for large squeezing.  A high degree of squeezing thus requires an exponentially large amount of irreversible work. Equations \eqref{10}, \eqref{14} and \eqref{15} reveal the intimate and simple relationship existing between the vacuum squeezing properties of the time--dependent harmonic oscillator on the one hand and its nonequilibrium thermodynamic properties on the other: the knowledge of the squeezing parameter  allows the determination of both the degree of nonadiabaticity of the frequency modulation and the amount of irreversible work produced. Conversely, the knowledge of the nonequilibrium thermodynamic state of the quantum oscillator  gives direct access to its vacuum squeezing. 

{\it Optimization of squeezing.} An important question  from a theoretical as well as practical point of view is the determination of a frequency protocol that leads to maximum squeezing for a prescribed maximal modulation amplitude. 
We use in the following optimal control theory  to answer this question for fixed values of the initial and final frequencies $\omega_0$ and $\omega_1$ of the harmonic oscillator. Optimal control theory (OCT) is a powerful mathematical optimization method based on the calculus of variations \cite{oct}. It allows to determine the function that minimizes a given cost functional in analogy to the familiar Euler--Lagrange equations of classical mechanics that minimize the action of a system. The results of a numerical implementation of OCT using Pontryagin's principle for  $\omega_0=1$ and $\omega_1=2$ are summarized in Figs.~1, 2 and 3. The oscillator is taken to be initially in the ground state and the cost functional to minimize is chosen as $1/\langle H(t)\rangle$, since the average energy is a monotonous function of the squeezing (see Eqs.~(\ref{eq9}) and (\ref{14})).
Figure 1 shows the optimal frequency modulation as a function of time; we observe that it consists of a regular sequence  of frequency jumps at which $\omega(t)$ abruptly switches from $\omega_0$ to $\omega_1$ and back. The corresponding squeezing parameter, depicted in Fig. 2, increases by discrete increments at each of the frequency jumps until the modulation is terminated. 

\begin{figure}
\center
\epsfig{file=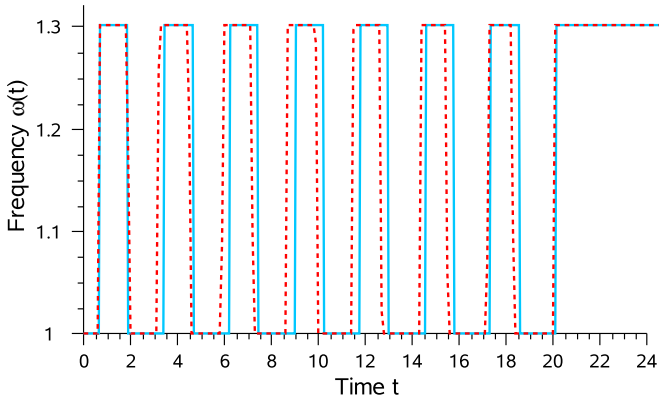, width=5.2cm, angle=270}
\caption{(color online) Frequency modulation of the harmonic oscillator as a function of time for fixed initial and final frequencies $\omega_0= 1$ and $\omega_1=1.3$. The dashed (red) line is the result obtained by optimal control theory and the solid (blue) line corresponds to the result of Ref.~\cite{jan92}.}
\label{fig:ramp}
\end{figure}

It is interesting to compare the above frequency protocol obtained with the help of OCT  to the one studied analytically by Janszky and Adam \cite{jan92}. Although the two protocols are not exactly identical, they lead to very similar squeezing results (see Figs.~1--3). Janszky and Adam considered a sequence of sudden frequency changes between $\omega_0$ and $\omega_1$, separated by some delay times $\tau_i$. They found that squeezing was strongest when these delay times were given by a quarter of the oscillation periods of the oscillator, that is, $\tau_0 = \pi/2\omega_0$ before a jump from $\omega_0$ to $\omega_1$ and  $\tau_1 = \pi/2\omega_1$ before a jump from $\omega_1$ to $\omega_0$. The latter condition exactly corresponds to the time needed to exchange position and momentum axes in phase space. The total squeezing generated after $n$ such cycles was shown to be
\begin{equation}
e^{2r}=(\omega_1/\omega_0)^n \ .
\end{equation}
\begin{figure}[t]
\center
\epsfig{file= 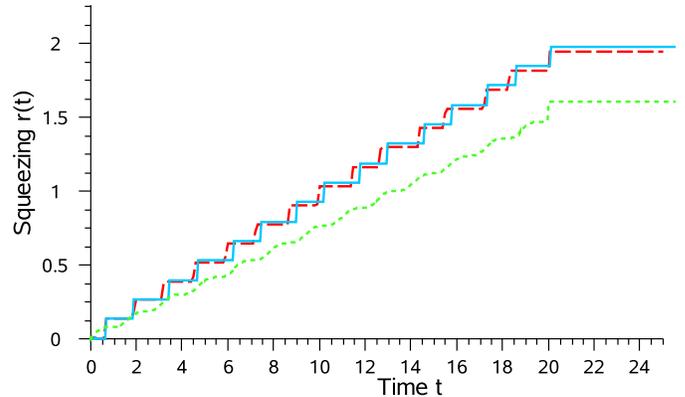, width=5.2cm, angle=270}
\caption{(color online) Vacuum squeezing of the harmonic oscillator generated by the frequency modulation shown in Fig.~1. The dashed (red) line is the result obtained by optimal control theory and the solid (blue) line corresponds to the result of Ref.~\cite{jan92}. The dotted (green) line shows the non--optimal squeezing produced by a periodic driving at  frequency $2 \omega_0$, $\omega(t)=\omega_0+(\omega_1-\omega_0)\sin(2\omega_0 t)/2$.}
\label{fig:sq}
\end{figure}

\noindent The degree of squeezing achieved by such a protocol therefore increases {\it exponentially} with the number of cycles. It then follows from Eq.~\eqref{14} that the nonadiabaticity parameter also grows exponentially with $n$  for large squeezing, $Q^*(t)\sim(\omega_1/\omega_0)^n/2$.
We have checked that the frequency modulation of Janszky and Adam is actually  a stable solution of the optimal control algorithm.

\begin{figure}[b]
\center
\epsfig{file= 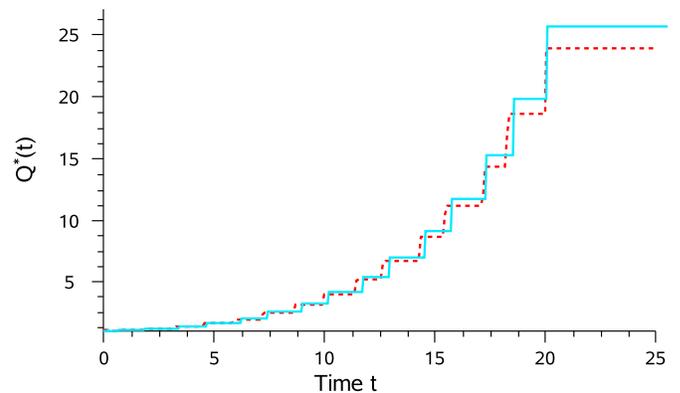, width=5.1cm, angle=270}
\caption{(color online) Nonadiabaticity parameter $Q^*(t)$, Eq.~(\ref{14}), generated by the frequency modulation shown in Fig.~1. The dashed (red) line is the result obtained by optimal control theory and the solid (blue) line corresponds to the result of Ref.~\cite{jan92}. }
\label{fig:ad}
\end{figure}
{\it Implementation in ions traps.}
A squeezed vacuum state of a quantum harmonic oscillator was created and observed experimentally using a single ion confined in  a Paul trap \cite{wineLett}. Paul traps are ultrastable rf traps that allow to prepare, manipulate and measure quantum states with high precision \cite{werth}. The first observation of quantum jumps was performed in a single ion trap \cite{ber86}. The frequency of a trap is determined by the external electrode voltages   and by  the size  of the trap. A voltage modulation  therefore directly leads to a modulation of the motional frequency. Since commercial electronic components can  achieve fast voltage switching rates, as  compared to the trap frequency,  tailored frequency variations can be  implemented. 

In the experiment \cite{wineLett}, the squeezed vacuum state of a harmonically confined $^9 \mbox{Be}^+$ ion was generated by  cooling the ion to its motional ground state using sideband cooling, and by irradiating it with two Raman beams with a frequency difference $2\omega_0$. The latter induces a parametric driving at frequency $2\omega_0$, which squeezes the ground state of the ion. 
The squeezed state was detected by laser--coupling  motional and electronic levels of the ion and observing the fluorescence signal of the ground state. The probability $P_g(t)$ that the ion remains in the {\it electronic} ground state  after a given coupling time $t$  depends on the level population of the {\it motional} degree of freedom. The probability distribution of the motional Fock state $P_n$ is a known function of the squeezing parameter and is given by the Fourier transform of the time signal $P_g(t)$ (mapped by sequential experimental runs with different final times).  The squeezing parameter $\beta= \exp(2 r)$ could  then  be deduced by  fitting the function $P_g(t)$, yielding  a value $\beta=40$. We can evaluate the  corresponding nonadiabaticity parameter  from Eq.~\eqref{14} to  be $Q^*(\tau)=20$. The squeezing protocol used in the experiment is not optimal and similar (or higher) squeezing values could be obtained in less time by employing the optimal squeezing modulation discussed in the previous section (see Fig.~\ref{fig:sq}).

It is worth noticing that the squeezing parameter can also be determined  from a measurement of the mean energy of the oscillator, by combining   Eqs.~(\ref{eq9}) and (\ref{14}).  The average energy  of the oscillator at a given time can be evaluated from the measured level population $P_n$ via the simple expression $\la H(t) \ra =\sum_n \hbar\omega_n(n+1/2)P_n$. By proceeding this way, the squeezing parameter  can hence be determined directly without doing any numerical fit.

{\it Conclusion.}
We have presented a relationship between the degree of nonadiabaticity, the irreversible work and the vacuum squeezing of frequency--modulated quantum harmonic oscillator. We have found that both the nonadiabaticity parameter $Q^*$ and the amount of irreversible work grow exponentially with large squeezing. We have moreover determined the optimal modulation that leads to   maximal squeezing using optimal control theory and found that the result is very similar to the protocol investigated analytically  by Jansky and Adam \cite{jan92}. We have in addition discussed the experimental implementation in single ion traps and proposed a new method to measure the degree of squeezing  and determine the nonadiabaticity parameter. 

This work was supported by  the
Emmy Noether Program of the DFG (Contract LU1382/1-1) and the
cluster of excellence Nanosystems Initiative Munich (NIM).

\end{document}